\documentclass{elsarticle}

\usepackage{amsmath}
\usepackage{url}
\usepackage{subcaption}
\usepackage{enumitem}
\usepackage{multirow}

\begin{document}


\title{Large-scale Biological Meta-database Management}

\author[uit]{Edvard Pedersen}
\ead{edvard.pedersen@uit.no}
\author[uit]{Lars Ailo Bongo}
\ead{larsab@cs.uit.no}

\address[uit]{Department of Computer Science and Center for Bioinformatics\\ University of Troms\o\\Troms\o, Norway}

\begin{abstract}
Up-to-date meta-databases are vital for the analysis of biological data. However, the current exponential increase in biological data leads to exponentially increasing meta-database sizes. Large-scale meta-database management is therefore an important challenge for production platforms providing services for biological data analysis. In particular, there is often a need either to run an analysis with a particular version of a meta-database, or to rerun an analysis with an updated meta-database. We present our GeStore approach for biological meta-database management. It provides efficient storage and runtime generation of specific meta-database versions, and efficient incremental updates for biological data analysis tools. The approach is transparent to the tools, and we provide a framework that makes it easy to integrate GeStore with biological data analysis frameworks. We present the GeStore system, an evaluation of the performance characteristics of the system, and an evaluation of the benefits for a biological data analysis workflow.
\end{abstract}


\maketitle


\newcommand{\ill}[1]{#1}
\newcommand{\exper}[1]{#1}
\newcommand{\figsize}[0]{0.7\textwidth}
\setlist{noitemsep}

\section{Introduction}
\label{introduction}

Recent advances in scientific instruments, such as next-generation sequencing machines, have the potential of producing data that provides views of biological processes at different resolutions and conditions, opening a new era in molecular biology and molecular medicine \cite{Schuster2008}. Many of the data analysis techniques developed for analyzing such biological data integrate data from many experiments with metadata from multiple knowledge bases. The information in the meta-databases \cite{Fernandez-Suarez2014} is essential for understanding the biological content of the experiment data. For example, the results of DNA sequencing may not become truly useful before the UniProtKB \cite{Magrane2011} meta-database is used to map sequence bases to genes, the gene expression results are compared to results from other experiments, and the differences in expression values have been  mapped to biological functions using the GO \cite{Ashburner2000} meta-database. 

\begin{figure}[h]
	\centering
    \includegraphics[width=0.5\textwidth]{\ill{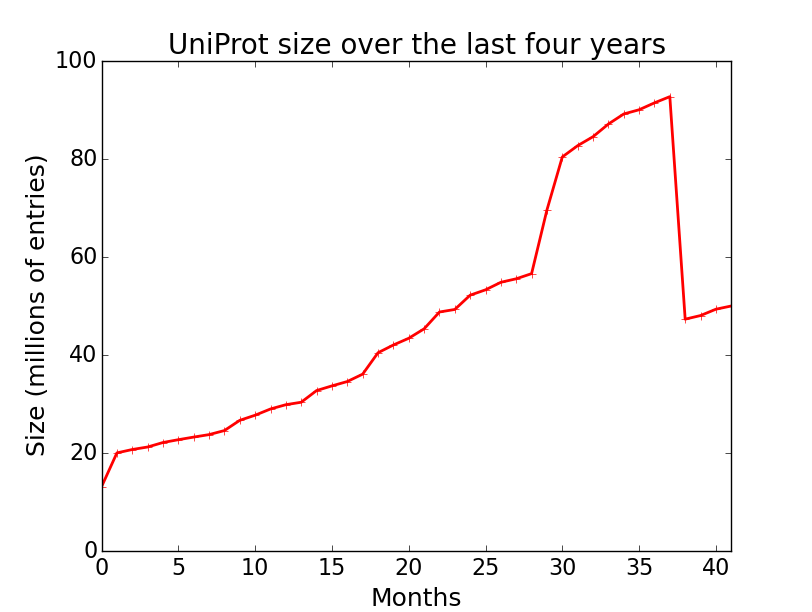}}
	\caption{Number of entries in UniProtKB from July 2011 to June 2015. The dip in early 2015 is due to removal of redundant proteomes \cite{uniprotRelNotes}.}
    \label{FIG1}
\end{figure}

The low cost of next-generation sequencing machines and other biotechnological instruments have caused an exponential growth of biological data \cite{Kahn2011}. Analysis of all this data produces many results, which are added to meta-databases such as UniProtKB. Such meta-databases are frequently updated and therefore growing rapidly (Figure \ref{FIG1}). For example, the June 2015 release of UniProtKB/TrEMBL contains 48.744.721 entries and is 137 GB in size. Compared to the previous May 2015 release, the number of entries increased by 5\%, and 45\% of the entries were updated. Each update may provide novel insights when reanalyzing old experiment data \cite{Sandberg2007}. Updating experiment data with new meta-data is especially important for servers that provide search analysis services based on integrated data analysis \cite{Benson2014}.

For many analyses, it is also important that a specific meta-database version is used. For example, it is common to compare analysis results against “gold standard” results that are calculated using a specific meta-database version.

There are four main requirements for an infrastructure system that maintains large-scale biological meta-databases. First, multiple versions of the meta-database must be maintained to ensure repeatability of the analysis. Such repeatability is a cornerstone in the scientific process, but has often been hard to achieve in bioinformatics \cite{Goecks2010}. Second, the system should enable efficient methods for integrating biological compendium with new or updated meta-data, since the computational cost of the integration can be orders of magnitude larger than the cost of producing the data \cite{Wilkening2009}. Third, the system must be transparent to data analysis tools since it is not practical to implement and maintain modified versions of the many analysis tools used in biological data analysis [10]. Fourth, the system must easily integrate with biological data analysis frameworks to ensure adaptation in production systems. 

Current popular biological data analysis frameworks such as Galaxy \cite{Goecks2010}, Taverna \cite{Oinn2004}, and Bioconductor \cite{Gentleman2004} do not satisfy the first two requirements, since the user manually maintains and specifies meta-data versions. In addition, meta-database updates typically require re-executing the analysis for each meta-data update. Such full updates increase the computational cost, often to the point where reanalysis is not done.

Incremental update systems \cite{Liu1998} for large-scale data \cite{Gunda:2010:NAM:1924943.1924949,Bhatotia2011,Peng2010,Popa2009,Logothetis2010,Bu2012} solve the first two requirements. These systems maintain several versions of the experiment data compendia and meta-databases, and greatly reduce the cost of reanalysis by using incremental updates that limits the computation to new and updated data. However, they do not provide a transparent approach for adding incremental updates to existing biological analysis workflows. Instead, they require either porting applications to a specific framework (such as Dryad \cite{Isard2007}, MapReduce \cite{Dean2008}, or Spark \cite{Zaharia2010a}) or implementing ad hoc scripts for input generation and output merging.

Data warehouse approaches for biological data, such as Turcu et al \cite{Turcu2008}, may provide incremental updates for specific tools, but do not easily allow adding new tools, nor integrating with biological data analysis frameworks.

We use the GeStore system \cite{pedersen2014transparent} for large-scale biological meta-database management. It satisfies all four requirements listed above. GeStore provides an efficient transparent file based approach for incremental updates. It uses HBase to implement distributed data structures with efficient compression for multiple versions of large meta-databases. It uses Hadoop MapReduce for scalable parallel generation of specific database versions and increments. The transparent approach enables easy integration of GeStore with data processing frameworks, and does not require any changes to data analysis tools.
Our contributions are threefold:


\begin{enumerate}
	\item We describe the design and implementation of a system for large-scale biological meta-database management.
	\item We demonstrate how the approach can be integrated with biological data analysis frameworks with minimal changes to the framework code, and no changes to data analysis tools.
	\item We present experimental evaluation of the performance, overhead and resource usage of the approach using a biological analysis workflows and real large-scale meta-databases.
\end{enumerate}

We find that large-scale biological meta-databases can be efficiently maintained using data-intensive computing systems, and that our approach can easily be integrated with biological data analysis frameworks.

\section{Background}
We provide a background describing biological data analysis implementation, configuration, and execution. Further examples can be found in \cite{Pedersen2015}, \cite{Bongo2014}.

\subsection{Data Analysis Workflows}

\begin{figure}[h]
	\centering
    \includegraphics[width=0.5\textwidth]{\ill{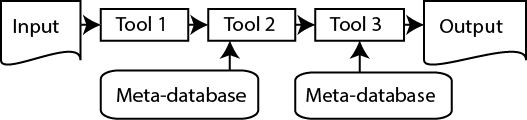}}
    \caption{A biological data analysis workflow}
    \label{FIG2}
\end{figure}

A computer system for analyzing biological data typically consist of four main components: input data, meta-data, a set of tools in a workflow, and finally output data for interactive analysis (Figure \ref{FIG2}). Biotechnology instruments such as short-read sequencing machines produce the input data. The data can also be downloaded from public repositories such as GEO \cite{Edgar2002} and ENA \cite{Leinonen2011}.  There are hundreds of meta-databases with human or machine curated meta-data extracted from the published literature and analysis of experimental data \cite{Fernandez-Suarez2014}. The datasets and databases range in size from megabytes to petabytes. 

A series of tools process the data in a pipeline where the output of one tool is the input to the next tool. The data transformations includes file conversion, data cleaning, normalization, and data integration. There are many libraries \cite{Goecks2010,Gentleman2004,Stajich2002} with hundreds of tools, ranging from small, user-created scripts to large, complex applications. A specific biological data analysis project often requires a deep workflow that combines many tools \cite{Diao2015}. 

\subsection{Workflow  Managers}
The analyst specifies, configures, and executes the workflow using a workflow manager. The workflow manager provides a way of specifying the tools and their parameters, management of data and meta-data, and execution of the tools. In addition, a workflow manager may enable data analysis reproducibility by maintaining provenance data such as the version and parameters of the executed tools. It may also maintain the content of input data files, meta-databases, output files, and possibly intermediate data.

A workflow manager may comprise of a set of scripts run in a specific platform, or a system that maps high-level workflow configuration to executable jobs for many platforms. There are also managers that provide a GUI for workflow configuration, and a backend that handles data management and tool execution.

\subsection{Hardware Platforms}
The workflow manager typically executes the workflow on a fat server, high performance computing clusters, or a data-intensive computing cluster.
Workflow managers such as Galaxy \cite{Goecks2010} are typically run on a single server. There are two main advantages. First, most biological analysis tools can be used unmodified. Second, it is not necessary to distribute and maintain tools on a cluster. The main disadvantage is the lack of scalability, both concerning dataset size and the number of concurrent users.

Script based workflow managers often execute the workflows on a high performance computing (HPC) cluster. Many biological data analysis tools can easily be run on such platforms by splitting the input (or meta-data) into many files that can be computed in parallel. The main advantage of using an HPC cluster is their availability. The main disadvantage is that the centralized storage system often becomes a bottleneck for production size datasets.

Systems such as Troilkatt \cite{Bongo2014,Diao2015} are designed to execute workflows on clusters built for data-intensive computing \cite{Shvachko2010}. Compared to HPC clusters these have storage distributed on the compute nodes, and data processing systems that utilize such distributed storage. The main advantage is improved performance and scalability for I/O bound jobs. The main disadvantage is that applications may need to be modified to utilize such a platform fully \cite{Diao2015,Pedersen2015,Bongo2014}.

\section{GeStore}
Our approach for large-scale meta-database management is based on the GeStore system \cite{pedersen2014transparent}. We built GeStore to enable transparent incremental computations for unmodified file-based data analysis workflows. Later we realized that these same approach and mechanisms are well suited, and needed, to efficiently maintain and generate specific versions of large meta-databases.

GeStore consists of a runtime system that provides a plugin framework for versioned input file generation and output file merging, a framework for parsing and detecting changes in files, distributed data storage, and parallel processing. GeStore uses HDFS \cite{Shvachko2010} and HBase \cite{Apache2012} to store meta-databases efficiently, and Hadoop MapReduce \cite{Dean2008} to generate a specific version of a large meta-database. In addition, GeStore provides library functions and tools to add incremental updates to workflows, and client applications to administer the data maintained by GeStore.

\begin{figure}[h]
	\centering
    \includegraphics[width=0.5\textwidth]{\ill{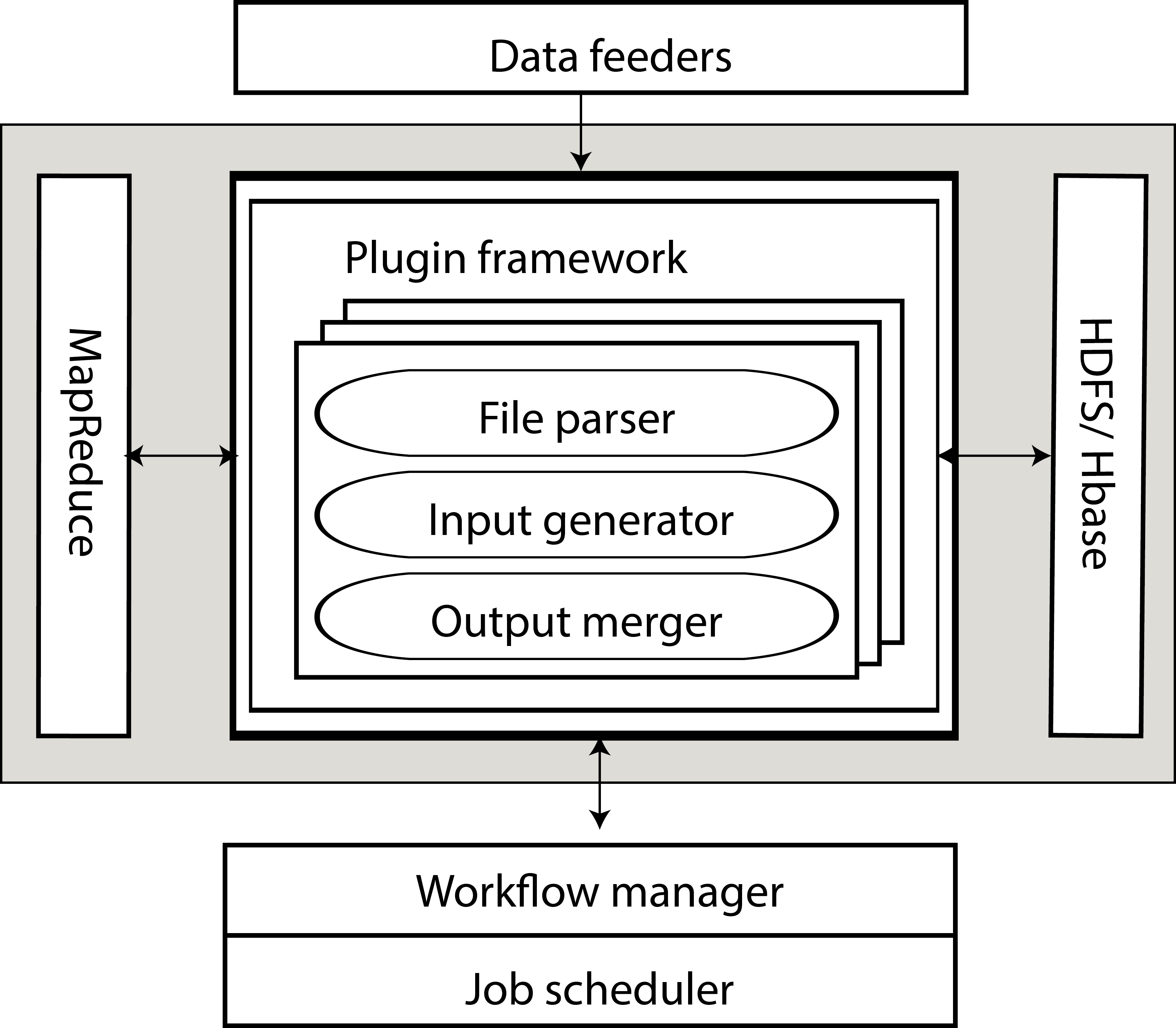}}
    \caption{GeStore architecture}
    \label{FIG3}
\end{figure}

GeStore exports two interfaces (Figure \ref{FIG3}). The first is for data feeders \cite{Shkapenyuk2011} that periodically download updated experiment and meta-data from public repositories and databases and add them to GeStore. The second is for workflow managers to get input and meta-data for a tool execution from GeStore, and merge output data with existing data after the tool is executed.

\subsection{File Based Incremental Updates}

GeStore uses a transparent file based approach to implement meta-database versioning and incremental updates by generating input and meta-data files for biological data analysis tools that only contain data for a specific period. For example, a meta-database for an incremental update may only contain the entries changed in the period. The tool will then be run, as normal, but it will typically produce a partial result in case of incremental updates. GeStore then merges the partial result with previously produced results.

We have chosen a file-based approach since many genomics applications use relatively few file formats. It is therefore feasible to implement parsers that support most file formats and therefore most tools. In addition, most file formats are simple and structured, which makes it easy to write parsers for each format. We also believe that many bioinformatics tools can use the file-based approach since bioinformatics applications are often parallelized using a data-parallel approach. Hence, a subset of the data can be computed separately, as in an incremental update.

One example of an analysis tool that is well suited for incremental updates is the widely used Basic Local Alignment Search Tool (BLAST \cite{blast}). It calculates a similarity score for all gene sequences in an input file by comparing each sequence to all sequences in the UniProtKB \cite{Magrane2011} meta-database. We can implement an incremental update of the results each time UniProtKB is updated by generating an incremental version of the database that only contains the entries that have changed since the last update.

The simplest approach to file generation is to compare all records in two versions of a file to find the new, deleted, and updated records. However, since most tools do not use all record fields, a naive diff will find too many changes. For example, BLAST results are not affected by the annotation record fields that are most frequently changed in the UniProtKB database. It is therefore necessary to write tool specific change detection that only detects changes in the significant fields. In addition, it may be necessary to handle new, updated and deleted records differently. For example, record deletions may require finding and discarding associated records in the output data.

The simplest approach to merge result files is to append the incremental updates to existing result files. However, some output record fields may contain values aggregated over the full dataset. For example, the BLAST output data contains a field, e-value, which is incorrect for incremental updates \cite{Turcu2008}. In such cases, the GeStore file merger must fix these values in the updated output files.

\subsection{Storage}
GeStore maintains versioned meta-database files, input files, and output files. GeStore uses the version information to generate incremental files, or a specific version of a meta-database. In addition, the version information is required to merge incremental update results with previously computed results. We use the Hadoop software stack for scalable storage and data processing.

There are two types of files maintained by GeStore: parsed and unparsed. Parsed files are stored as HBase tables, while unparsed files are stored unmodified in HDFS. The unparsed files are files that are difficult to parse and/or do not need incremental updates, such as files that are always completely updated, or not at all. 

For file types that have a parser implemented, GeStore splits the data into entries and entry fields. The entries are stored as rows in HBase, and the fields as columns. All entries are stored in the same HBase column family. The only required columns in the schema is a unique ID for each row that the plugin uses to generate a row key, and an EXISTS column that specifies  if a given entry exists in the current version of the meta-database.  The remaining columns are file-format specific. HBase is designed such that new columns can easily be added to a table. GeStore uses this flexibility to enable reuse an old HBase table even if the file format or the parser code is changed.

We use the HBase timestamp mechanism to manage meta-database versions. The timestamp either represent the file generation date, release date or version of the meta-database. By storing updated data in timestamped HBase cells, we can efficiently compress  database versions using delta compression. In addition, HBase tables are compressed using the Snappy algorithm.

\subsection{Processing}
\label{processing}
GeStore provides four operations on the meta-database tables: create, update, get version, and get increment.

To create a table for a new meta-database, GeStore creates an empty HBase table, with one empty column family. Additional columns are added later for the meta-database fields as described in section \ref{pluginframework}.

To update an existing table with new meta-data, GeStore first finds the correct table to use, then updates or adds new rows using a parallel job that executes the parser for the specific meta-database. Each entry in the new meta-database is compared to the entry in the previous version by comparing the corresponding HBase row. If there are no changes no updates are made, except to the EXISTS field. If one or more fields have changed, the column in the row are updated with the new data for the field and with the current timestamp. If new fields are added, a new column is added to the table with the new data and timestamp. If the row is deleted the EXISTS column is not updated. 

To generate an incremental update a parallel job is executed that scans the table for the timeframe $T_{last run} - T_{current time}$. For each record in the scan, the fields that are relevant for the specified output are selected. If there are updates to one or more of these fields, and the record has a current EXISTS field, the relevant record fields are written to a file on HDFS.

To get a specific version of the meta-database we use the above approach with the timeframe $T_{first} - T_{specified}$.

\subsection{Meta-Database Caching and Internal Data Structures}

GeStore implements a cache of previously generated meta-database files, since many workflows can share these. The files are stored in HDFS, and the filename is used to store information about how the files were generated.  We store both big (multi-gigabyte) and small (megabyte) files, since the overhead of generating both is large.
When a workflow manager requests a meta-database, GeStore uses the generated filename to lookup in the cache. Matching files are returned from the HDFS cache. Otherwise, a new file is generated and cached. GeStore does not limit the cache size, but the oldest files in the cache can automatically be deleted by e.g. a cron job.

In addition, GeStore maintains information about each update for each file it maintains, and the files accessed by each workflow tool execution. This table is used to identify which files a workflow used, when they were used, and how they were generated.  

\subsection{Plugin Framework}
\label{pluginframework}

To use GeStore to maintain the meta-data used by pipeline tool, the workflow maintainer must implement: (i) a parser for each file type used by the tool, (ii) tool-specific file generator, and (iii) tool-specific incremental output file merger (if incremental updates will be used). In GeStore, these are implemented as a plugin, and managed by the GeStore plugin framework. The framework is invoked by a workflow manager or data feeder that respectively calls one of the two exported functions: generateFiles or mergeFiles. The plugin framework uses MapReduce jobs to do the processing required to add data to the system and retrieve it, as well as doing change detection, data verification and merging. MapReduce is used since the files can be very large, and hence efficient parallel processing is needed.

The plugins parse and store data, while the (unmodified) tool does the data analysis. It is easy to implement a plugin. Typically, only a few tens of lines of code must be written, since many plugins can reuse parsers and file mergers written for other plugins. In addition, the framework provides a library of parsers for known file formats, and libraries for parsing, change detection, and merging of files. GeStore also takes care of efficient data storage, low overhead file parsing, file generation, and merging. 

\subsubsection{File Parser}
The file parser must determine the structure of input files and meta-database files used by a tool. Only one parser must be implemented for each file format, so it is likely that parsers already exists for the file formats used by a tool. The parser must also convert the file data into the HBase table format used by GeStore. Most biological data is in a table format so it is usually easy to implement a parser. 

The file parser interface consists of six methods that must be implemented. These: (i) define the start and end of an entry in the file using regular expressions, (ii) split an entry into columns, (iii) compare two versions of an entry, (iv) check if an entry contains all elements required by the tool, (v) generate a HBase Put object, and (vi) generate output in other formats. 

\subsubsection{File Generator}
The file generator class is responsible for generating the input and meta-data files used by a tool. It must detect changes in input data and meta-data. The change detection can be course grained, where the contents of an entire file is compared, or fine grained where individual records are compared. For the latter, the change detection may take into account the structure of the file. In particular, the change detection is efficient and easy to implement if the data is stored in HBase tables as described above. 

The file generator requires implementing one method that specifies the parsers to use for each file format, and the fields to write to the input file using the associated file parser.

\subsubsection{Output Merger}
\label{merge}

The output merger is responsible for merging the results of an incremental computation with previously generated output stored in GeStore. GeStore executes output merging similarly to meta-database updates. However, the merge is application-dependent, and hence requires application specific knowledge to understand how an incremental computation may influence the results and how to fix any resulting errors. To fix errors a tool-specific method must be implemented in the plugin.

\subsection{Workflow Manager Integration}

The workflow manager must call GeStore to execute the plugin for a tool. GeStore provides a minimal interface with three functions that the workflow calls to request: a specific meta-database version, one or more incremental update input files, or to merge the partial results with previously produced results. In addition, GeStore provides a semi-POSIX file system interface implemented as a Java library for low-level integration with workflow managers. To use it, the workflow manager code is modified to replace calls to for example HDFS to the respective GeStore functions.

\section{META-pipe Plugins}

In this section, we describe how we added incremental updates to the META-pipe metagenomics analysis workflow. Integration approaches for two additional workflows and workflow managers are described in \cite{Pedersen2015}.

\subsection{META-pipe}
\label{meta-pipe}
META-pipe is a DNA sequence analysis workflow used by our biology collaborators to find novel commercially exploitable enzymes from marine microbial communities. These communities are not well explored, so a meta-database update can significantly change the analysis results. META-pipe takes as input assembled reads from environmental marine samples. The analyst uses tools such as METArep \cite{Goll2010a} to visualize and explore the workflow output results. 

We use a smaller version of the full META-pipe workflow for our application benchmarks. We do not include the final annotation step, since a bug in one of the tools caused the execution time to be much higher when GeStore was \emph{not} used. The version of META-pipe we use comprise the following tools:
\begin{itemize}
	\item MetaGeneAnnotator (MGA) \cite{Noguchi2008}: predicts genes in metagenomic sequences by searching for start sites and ribosomal binding sites. 
	\item MGA-exporter: a Perl script that converts the MGA output files to the format used by the next stage.
	\item FileScheduler: a python script that partitions and distributes the input data to the compute nodes in a cluster.
	\item Protein BLAST (BLASTP) \cite{blast}: maps sequences to meta-data found in UniProtKB \cite{Magrane2011}.
\end{itemize}

\subsection{Plugins}
To add incremental updates to META-pipe we implemented parsers for the four file formats used by the workflow: FASTA, UniprotKB meta-data, BLAST output, and MGA output. In addition the BLAST tool plugin corrects incremental e-values as discussed in \cite{Turcu2008} during a merge. The file format plugins were 580 lines of Java code, and the tool plugins were 260 lines of Java code. 

\subsection{Workflow Manager Integration}

The Meta-pipe workflow manager, called GePan, uses the key-value store interface exported by GeStore. We chose to focus on GePan, since it is the workflow manager that we have deployed on our production system. We have modified the GePan script generation to replace file copy operations in the job scripts with GeStore calls. GePan sets the GeStore arguments at runtime, and specifies the incremental files to generate and meta-database versions to use. 

The development effort for the integration was high. All file accesses are from scripts generated by GePan, hence GePan must be modified to replace these file accesses with GeStore calls. In total, we added about 300 lines of code to the 14.000 line GePan codebase. However, we did not modify any of the META-pipe tools.

\section{Evaluation}

In our previous work \cite{pedersen2014transparent}, we evaluated the overhead and performance improvements of GeStore for incremental updates for biological data analysis pipelines. In this paper, we focus on issues related to deployment of GeStore in a production system. In particular, we want to answer the following two questions: (i) What are the performance and resource usage characteristics of GeStore? (ii) How does the overhead of GeStore compare to alternative approaches for biological meta-data management?

We evaluate the first question to understand how to deploy GeStore in a production system, including identifying areas for optimization, understanding the scalability of the GeStore operations, and how GeStore perturbs other applications running concurrently on the same system. The answer to the second question demonstrates the usefulness of GeStore meta-data management for biological data analysis.

\subsection{Methodology}

We characterize GeStore performance and resource usage using benchmarks for each of the GeStore operations. We use the Meta-pipe workflow (section \ref{meta-pipe}) as an application benchmark. In addition, we have implemented ad hoc tools for Meta-pipe meta-data management.

We report the average execution time of the benchmarks. Each experiment is repeated 3 times. The standard deviation is less than 5\% of runtime for all experiments and is therefore not reported. 

We use the Ganglia Monitoring System \cite{Massie2004} to measure CPU load, memory usage, network traffic, and disk I/O during benchmark execution.
The experiments were run on a 10-node cluster. It has one front-end node with and NFS server, one node with HDFS namenode and HBase master server, and eight HDFS/HBase/MapReduce/Cassandra data/compute nodes. Each node has 32 GB of DRAM, a 4-core Intel Xeon E5-1620 CPU with two-way hyper-threading, 4 TB local disk, and a 2 TB disk used for NFS. The cluster has a 1-gigabit Ethernet interconnect. We assume the cluster size and configuration is realistic for a small cluster in a production environment.

The software used is Oracle Java 1.7.0, Cloudera 4.6.0 (HBase 0.94.6, HDFS 2.0.0, MapReduce 2.0.0, ZooKeeper 3.4.5). In addition, the UniProtKB plugin uses formatdb 2.2.25, which is part of the legacy BLAST package.

HBase is configured to use a heap size of 4 GB for the master server and 12 GB for the region servers. HDFS is configured with a replication factor of three, block size of 128 MB, and heap size of 1 GB for the NameNode and DataNodes. HBase is configured to use Snappy compression, and has a maximum of 32 write-ahead log files. Client scan caching is set to 100.

The data used are the latest UniProt meta-database\footnote{Available at \url{http://www.uniprot.org}} (versions are specified below), and a metagenomic sample from the Yellowstone National Park \cite{Bhaya2007}\footnote{Data available at \url{http://metagenomics.anl.gov/linkin.cgi?metagenome=4443749.3}}. GeStore version 0.2 is used\footnote{Available at \url{http://github.com/EdvardPedersen/GeStore}}, as well as a modified version of GePan\footnote{Available at \url{http://github.com/EdvardPedersen/GeStoreGePan}}.

\subsection{Add and Update Meta-Databases}

We first measure the time and resource usage of adding a new meta-database to GeStore, and for updating an existing database. We assume new meta-databases are rarely added, and that meta-databases are updated at most weekly. Both operations are therefore background operations, and we are therefore primarily interested in their resource usage. Also note, that a GeStore merge is executed similarly to an update.

We download the September 2014 release of UniProtKB (41 GB, gzip compressed), and decompress it on the frontend (231 GB). We measure the time of copying the files to HDFS, and then running a MapReduce job that reads and parses the HDFS files, and puts the 84.5 million parsed entries to an empty HBase table. To update the meta-database, we updated it with entries that were updated in the October release (37 out of 87 million entries). The MapReduce job for the update reads and parses the HDFS file, reads old entries by scanning the HBase table, compares each old and new entry, and puts updated entries to HBase.

\begin{table}[ht]
    \centering
    \small
	\begin{tabular}{|l|l|}
		\hline
		Add 2014\_09 UniProtKB & 182 minutes \\ \hline
		Update to 2014\_10 UniProtKB & 144 minutes \\ \hline
		Retrieve UniProtKB& 36 minutes \\ \hline
		Retrieve cached UniProtKB & 12 minutes \\ \hline
		Retrieve incremental UniProtKB & 5 minutes \\ \hline		
		Retrieve cached incremental UniProtKB & 26 seconds \\ \hline
	\end{tabular}
	\caption{GeStore add, update, and retrieve execution times.}
	\label{TAB1}
\end{table}

The time to add the UniProtKB meta-database to GeStore is 182 minutes (Table \ref{TAB1}). Updating the meta-database is 21\% faster (144 minutes). In addition, the time to download and decompress the database are respectively 52 and 33 minutes. We believe the update operation is faster, even if it requires scanning 84.5 rows from HBase for two reasons: First, there are fewer entries put to the HBase table. Second, the meta-database is (mostly) cached in memory on the HBase region servers and the updates are more evenly distributed among the cluster nodes. 

The resource usage of the add and update operations are similar. Both use 80\% of the maximum aggregated bandwidth of the interconnect, and have about 50\% CPU utilization over all eight cores. We therefore believe the performance is limited by the HBase (and HDFS) operations. Performance may be improved by better tuning of these operations, for example by optimizing client-side buffering of put operations. Performance will also improve by disabling or relaxing the write-ahead-log (WAL) for HBase. But this increases the chance of table corruption. The scalability of the add and update operations is similar to the scalability of HBase read and write intensive jobs. The results also show that there are CPU cycles available on the cluster for a computation-intensive job run concurrently with these operations. 

\subsection{Retrieve Meta-Databases}
\label{retrieve}

We evaluate the overhead and resource usage of retrieving an existing meta-database from GeStore, and saving it on a local file system to be used by a non-distributed analysis tool. This is a common operation for analysis pipelines with legacy analysis tools. Since the retrieve contributes to pipeline execution time, it should have a low overhead.

We first measured the time to retrieve the November 2014 version of the UniProtKB meta-database (240 GB uncompressed, 89 million entries) from GeStore (table \ref{TAB1}). GeStore will first run a MapReduce job with a large number of map tasks that retrieve the relevant fields for each entry (HBase row), and a single reduce task that writes the output to a single file. The total time to retrieve the meta-database is 36 minutes, in which the mappers are done after 14 minutes, and the reducer runs an additional 17 minutes. The overhead is acceptable since legacy analysis tools often have a step that must be run sequentially. For example, to convert the retrieved UniProtKB meta-database to a BLAST-compatible format using the formatdb tool (resulting file is 32 GB), requires an additional 36 minutes, with an additional 8 minutes to copy the file to HDFS. In addition, the total pipeline execution is often several hours or more.

Second, we retrieved an incremental version of UniProtKB that contains the entries updated between the September 2014 and October 2014 releases (in total 2.7 million entries, resulting in a file size of 1 GB). The incremental database is generated in 9 minutes. The speedup is due to the much smaller resulting file size and hence reducer execution time.

The cached version of the full UniProtKB takes 12 minutes to retrieve, and the incremental version takes 26 seconds. We believe many pipeline executions can use the cached meta-databases in a production system.

\begin{figure}[ht]
	\centering
	\begin{subfigure}[b]{0.49\textwidth}
		\includegraphics[width=\textwidth]{\exper{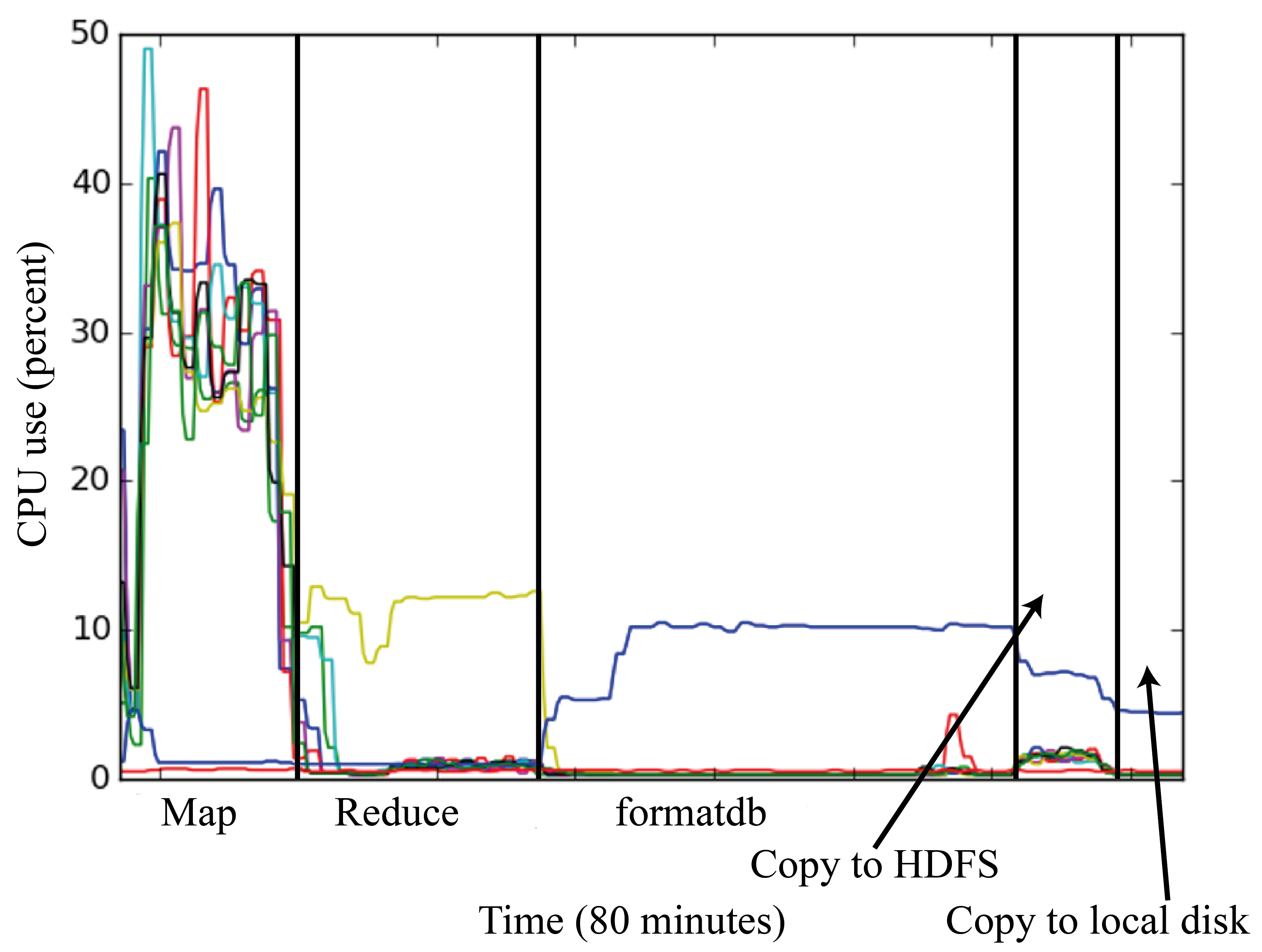}}
		\caption{Full meta-database}
		\label{fullCpuFig}
	\end{subfigure}
	\begin{subfigure}[b]{0.49\textwidth}
		\includegraphics[width=\textwidth]{\exper{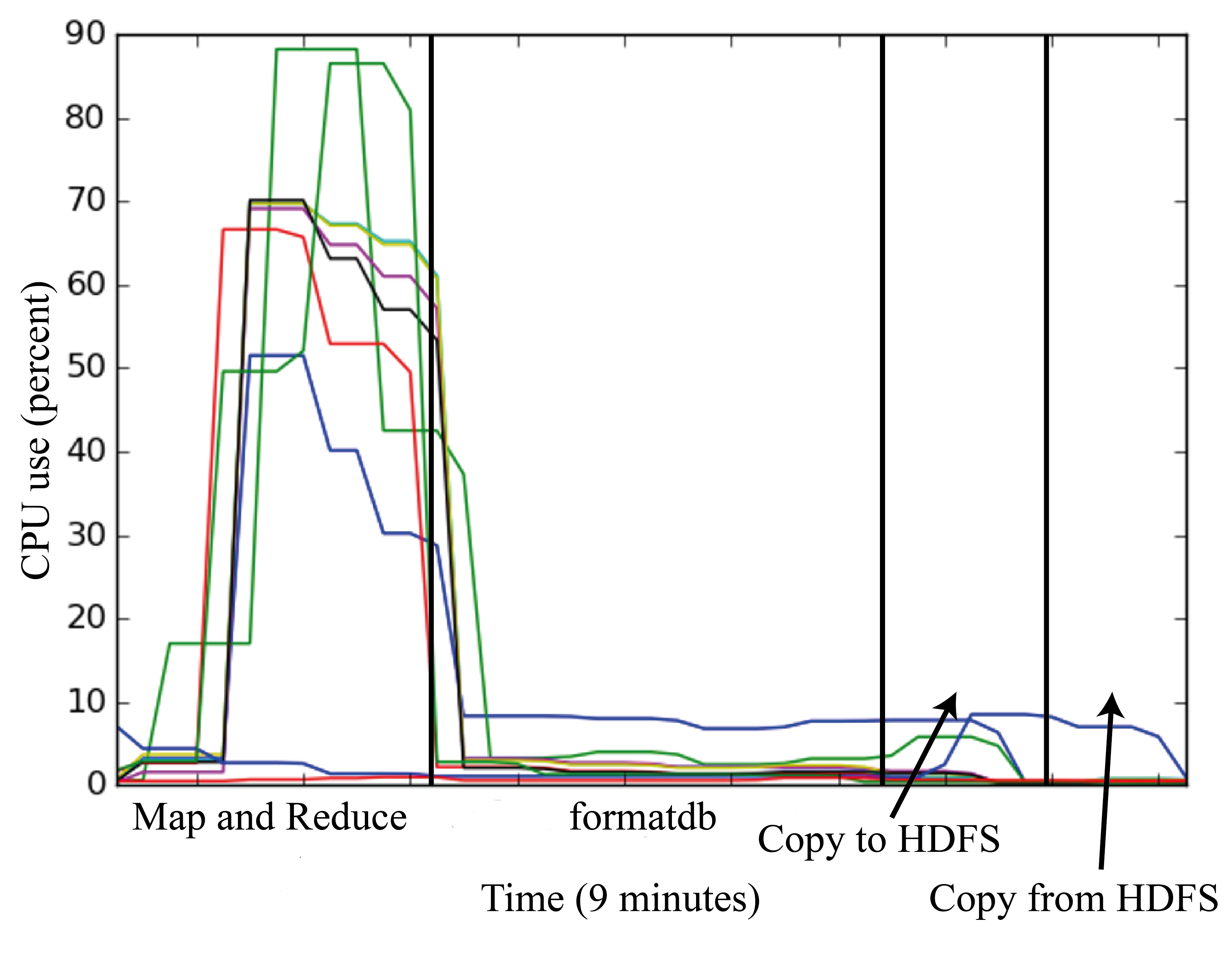}}
		\caption{Incremental meta-database}
		\label{incrCpuFig}
	\end{subfigure}
	\caption{CPU use when generating meta-databases, showing the different stages of processing}
	\label{CPUFIG}
\end{figure}

The maximum network utilization is about 20\%. The CPU utilization differs between the full and incremental update (Figure \ref{CPUFIG}) since there is more processing per byte of data in the incremental case. Figure \ref{CPUFIG} also shows how the legacy tools (formatdb, copy to HDFS and copy to local disk), as well as the single reducer stage, dominate the execution time. These execute sequentially and hence limit scalability. In the next section, we evaluate retrieve for tools that do not have this limitation. The incremental update utilizes most of the cluster resources and is therefore not suited to execute concurrently with another job. However, the execution time is short. The longer executing retrieve for a specific version is well suited to overlap with another job, especially during reduce.

\subsection{Retrieve and Split Meta-Database}

In the experiments in the previous sections, the performance of the retrieve operation was limited by the need to format and write the meta-database to a NFS or local file system. In this section, we measure meta-database retrieve time for  biological data analysis tools that can utilize the high aggregate disk bandwidth on a data-intensive computing platform by reading splits directly from a distributed file system and computing on these in parallel.

We initialized an HBase table with a 50 GB FASTA file with 150 million entries (sequences). The file is generated as in section \ref{retrieve}, but we do not run the formatdb tool, and the output is split among 20 reducers that each write their split directly to the HDFS cache.

\begin{table}[ht]
	\centering
	\small
    \begin{tabular}{|l|l|}
        \hline
        Retrieve FASTA & 55 minutes \\ \hline
        Retrieve cached FASTA & 10 minutes \\ \hline
        Retrieve and split FASTA & 9 minutes \\ \hline
        Get HDFS path of cached FASTA file & 2 seconds\\ \hline
    \end{tabular}
    \caption{Execution time for retrieve and split for the FASTA meta-database.}
    \label{TAB3}
\end{table}

The execution time is reduced from 55 to 9 minutes (Table \ref{TAB3}). The cached version can be read directly from HDFS, and therefore incurs no overhead. Since there is no bottleneck due to a single reducer, the retrieve scales well. The mappers utilize the CPUs up to around 70\%, and the mappers utilize the network close to 80\%.

\subsection{Storage System Comparison}

Above we found that the performance of the GeStore operations is largely dependent on the underlying distributed storage system. In this section, we compare the performance of the HBase system used by GeStore to the Cassandra system \cite{Cassandra}. We use a microbenchmark that generates a workload with the characteristic of GeStore operations. 

The microbenchmark uses the SPROT part of UniProtKB. We measure the time it takes to add this data to respectively HBase and Cassandra, and then retrieve it. We use a parallel implementation where 8 compute nodes each insert 1/8th of the data. The rows are prefixes with the node hostname, such that these later can be retrieved by a process running on the node. We use a replication factor of 3, so most requests can be served locally. But there are still some remote data accesses. 

Our results shows that for retrieves HBase and Cassandra have similar performance (respectively 9.6 sec and 9 sec for 547085 entries). Cassandra is however much faster than HBase for inserts (58 vs. 161 seconds). However, although we could implement the GeStore operations in Cassandra, we prefer HBase due to our experience with the Hadoop system, and the close integration with MapReduce. In addition, for GeStore retrieve performance is more important than insert performance as discussed above.

\subsection{Space Usage}

In this section, we quantify the storage overhead of GeStore. We measure the storage space used by the UniProtKB meta-database when compressed using gzip, uncompressed, and in when stored in HBase with delta and Snappy compression and 3x replication.

The disk space used by GeStore is comparable to storage of the compressed meta-databases (Table \ref{sizetab}). The overhead for a single version in HBase is high. With additional versions, the space usage first decreases, then it increases roughly with the size of the compressed version of each new version of UniProtKB. We believe the storage overhead is acceptable, since low-cost distributed storage is used.

\begin{table}[ht]
	\centering
	\small
	\begin{tabular}{|l|l|l|l|}
		\hline
		Versions & Compressed on disk & Uncompressed on disk & In HBase \\ \hline
		2014-03            &  23 GB & 133 GB & 306 GB \\ \hline
		2014-03 to 2014-04 &  47 GB & 268 GB & 240 GB \\ \hline
		2014-03 to 2014-05 &  71 GB & 405 GB & 210 GB \\ \hline
		2014-03 to 2014-06 &  99 GB & 568 GB & 234 GB \\ \hline
		2014-03 to 2014-07 & 133 GB & 757 GB & 273 GB \\ \hline
	\end{tabular}
	\caption{Aggregate size of UniProtKB on disk and in HBase using snappy and delta compression with a replication factor of three.}
	\label{sizetab}
\end{table}

\subsection{Application Benchmarks}

To evaluate the benefits of GeStore incremental updates for a real-world biological analysis workflow, we used Meta-pipe (described in section \ref{meta-pipe}). Meta-pipe scales linearly with respect to input data and meta-database sizes. Hence, a reduction in meta-database size will reduce the execution time approximately linearly with the size of the meta-databases used. GeStore will therefore improve performance if the GeStore operations have lower overhead than the reduction due to the smaller incremental meta-database size. We did a similar experiment in \cite{pedersen2014transparent}, but the results presented here are for the updated code and with (at the time we did the experiments) the latest UniProtKB meta-databases.

We measure the execution time of Meta-pipe using a small (15 MB) input file. Only the MGA, MGA exporter, fileScheduler and BLAST were run. We use the 2014\_09, 2014\_10 and 2015\_01 versions of UniProtKB. With a larger input file, the overhead of GeStore compared to workflow execution time will be even lower.

\begin{table}[ht]
	\centering
	\small
    \begin{tabular}{|l|l|}
        \hline
        Full update without GeStore & 833 minutes \\ \hline
        Full update with GeStore & 965 minutes \\ \hline
        Full update with GeStore, cached DB & 859 minutes \\ \hline
        1-month incremental update & 61 minutes \\ \hline
        4-month incremental update & 99 minutes \\ \hline
    \end{tabular}
    \caption{Application benchmarks for Meta-Pipe}
    \label{TAB4}
\end{table}

GeStore adds an overhead of 132 minutes when generating a meta-database, and 26 minutes when the meta-database is cached (Table \ref{TAB4}). Incremental updates are done in 61 minutes for the 1-month update, and 99 minutes for the 4-month update. A 1-month incremental update has a 13-fold speedup compared to a full re-analysis. These results show that incremental updates through GeStore can provide large benefits to biological analysis workflows.

\subsection{Comparison to Ad Hoc Approaches}

In this section, we compare GeStore performance to the non-distributed ad hoc meta-data management approach used by most biological data analysis pipelines. We will focus on the BLAST and \emph{annotation} stage of Meta-pipe. These stages requires a BLAST database and annotation database on each of the compute nodes for parallel execution. It is therefore necessary to generate these meta-databases and then replicating them to all nodes. We cannot store the databases on NFS, since the many small file reads by the pipeline significantly increases pipeline execution time. We have implement a Python script\footnote{Available at \url{http://github.com/EdvardPedersen/SimpleMetaManager}} that extracts the FASTA version of UniProtKB (release 2014\_11), uses formatdb to generate the BLAST database, and puts the database into a SQLite database used by the annotation tool. Finally, we distribute the resulting files to the computation nodes using rsync with deflate compression enabled. 

The creation of the SQLite database is comparable to a GeStore add, since they both make the database accessible from all nodes. The GeStore add is 191 minutes, and the ad hoc script execution time is 315 minutes. Write FASTA, formatdb and copying the BLAST database to the nodes is comparable to a GeStore retrieve, which takes 80 minutes for GeStore, and 206 minutes for the ad hoc script. We have not optimized the ad hoc script. However, we believe such unoptimized scripts are common in biological data analysis. In addition, optimizations to the GeStore operations will benefit all analysis tools using GeStore, while optimization of an ad hoc script typically only benefits a single pipeline.

We have not found a realistic use case for an ad hoc implementation of a meta-database update and generation of an incremental meta-database. For these, existing biological data analysis frameworks typically require generating a new database and then re-executing the full pipeline.

\begin{table}[ht]
	\centering
	\small
	\begin{tabular}{l|l|l|}
		\cline{2-3} 									& Ad hoc Script 		& GeStore \\ \hline
		\multicolumn{1}{|l|}{Write FASTA} 				& 63 minutes 			&  \\ \cline{1-2}
		\multicolumn{1}{|l|}{Formatdb}			 		& 34 minutes 			&  \\ \cline{1-2}
		\multicolumn{1}{|l|}{Copy BLAST DB to nodes} 	& 109 minutes 			& \\ \hline
		\multicolumn{1}{|l|}{\emph{Total}} 				& \emph{206 minutes} 	& \emph{80 minutes} \\ \hline \hline
		\multicolumn{1}{|l|}{Create SQLite DB} 			& 142 minutes 			& \\ \cline{1-2}
		\multicolumn{1}{|l|}{Copy SQLite DB to nodes} 	& 173 minutes 			& \\ \hline
		\multicolumn{1}{|l|}{\emph{Total}} 				& \emph{315 minutes} 	& \emph{191 minutes} \\ \hline
	\end{tabular}
	\caption{Ad hoc scripts vs. corresponding GeStore operations.}
	\label{TAB5}
\end{table}

\section{Related Work}

Management of large-scale data relies on a robust storage system, we have chosen to use HDFS \cite{Shvachko2010} and HBase \cite{Apache2012} due to the close relationship with the Hadoop framework, but other alternatives include Cassandra \cite{Cassandra}, MongoDB \cite{MongoDB} and BlobSeer\cite{blobseer}. In addition, we need to perform some processing on the meta-databases to implement the features of GeStore, we have chosen MapReduce \cite{Dean2008}, but similar systems such as Spark \cite{Zaharia2010a}, Nectar \cite{Gunda:2010:NAM:1924943.1924949} and Dryad \cite{Isard2007}, provide many of the same benefits. A more traditional system with e.g. pNFS \cite{pnfs} and MySQL \cite{mysql} are also an option, but would also require a distributed processing framework to fully utilize the cluster.


These processing systems have also been extended to support incremental processing, through systems like Incoop \cite{Bhatotia2011}, Percolator \cite{Peng2010}, and Marimba \cite{Schildgen2014}. In addition, data aggregation systems such as in \cite{ddas} extend the processing systems to support processing and management of general data types. We have used many ideas from these systems when designing GeStore.


Simple change detection is supported by tools such as UNIX diff, delta encoding compression systems \cite{Douglis2003}, and version management systems such as CVS. However, the change detection in these do not take into account the complex inter-file relationships found in genomic datasets.
GeStore extends the work in \cite{Turcu2008} by providing a framework and libraries to implement the necessary pre and post processing of data moved between a data warehouse and genomic analysis tools. This makes it easier to add additional support for additional genomic analysis tools as we have demonstrated by implementing incremental updates for a complete metagenomics analysis workflow.

\section{Conclusion}

We proposed an approach for efficient management of large-scale biological meta-databases. The approach is designed for production systems where biological analysis workflows are periodically executed to analyze large-scale datasets, often by updating existing analysis results with new meta-data. We presented the design and implementation of the GeStore system, including a framework for implementing plugins that enable transparent incremental updates. We demonstrated the feasibility of our approach and provided an experimental evaluation of our system for meta-database management using a real metagenomics analysis workflow and real data. Our findings show that large-scale biological meta-databases can be efficiently maintained using data-intensive computing systems, and that our approach can easily be integrated with biological data analysis frameworks, replacing ad hoc solutions. We have also characterized the performance and resource usage of meta-database management operations and provided insight into how GeStore can be deployed on a production system.

We plan to deploy GeStore on our production systems and use it to produce data for tools such as META-pipe.
 
The GeStore source code, a users guide describing the GeStore API, a tutorial for setting up and using GeStore, as well as a Vagrantfile for automatically installing the Hadoop stack and GeStore on a virtual machine, are all available at: \url{http://github.com/EdvardPedersen/GeStore}

\section*{Acknowledgements}
Thanks to Espen Robertsen and Tim Kahlke for help with the GePan workflow, Jon Ivar Kristiansen for maintaining our cluster, Nils Peder Willassen for his biological insights, Martin Ernstsen for his input on tuning HBase, Bj{\o}rn Fjukstad and Einar Holsb{\o} for their insightful comments. We would also like to thank the reviewers for their comments and suggestions.

\section*{References}

\setlength{\bibsep}{0pt plus 0.3ex}

\bibliographystyle{elsarticle-num}
{\footnotesize
\bibliography{curated}}

\end{document}